\documentclass[letterpaper,11pt]{article} %
\usepackage{url}  %
\usepackage{graphicx}  %
\usepackage[ascii]{inputenc} 
\usepackage{amsthm}
\usepackage{amsmath} %

\newcommand{\bigo}{{\protect\cal O}}

\newcommand{\p}{\ensuremath{\mathrm{P}}}
\newcommand{\np}{\ensuremath{\mathrm{NP}}}
\newcommand{\pp}{\ensuremath{\mathrm{PP}}}
\newcommand{\sharpp}{\ensuremath{\mathrm{\#P}}}
\newcommand{\ph}{\ensuremath{\mathrm{PH}}}
\newcommand{\pspace}{\ensuremath{\mathrm{PSPACE}}}
\newcommand{\boldthetatwo}{{\mbox{\boldmath\ensuremath{\Theta_2^p}}}}
\newcommand{\thetatwo}{\ensuremath{\Theta_2^p}}

\newcommand{\conp}{\ensuremath{\mathrm{coNP}}}

\newcount\hour  \newcount\minutes  \hour=\time  \divide\hour by 60
\minutes=\hour  \multiply\minutes by -60  \advance\minutes by \time
\def\mmmddyyyy{\ifcase\month\or Jan\or Feb\or Mar\or Apr\or May\or Jun\or Jul\or Aug\or Sep\or Oct\or Nov\or Dec\fi \space\number\day, \number\year}
\def\hhmm{\ifnum\hour<10 0\fi\number\hour :%
  \ifnum\minutes<10 0\fi\number\minutes}

\let\shortcite\cite

\begin{document}
\sloppy

\author{%
        Lane A. Hemaspaandra\\
        Department of Computer Science \\
        University of Rochester \\
        Rochester, NY 14627, USA}

\date{October 30, 2017; revised November 21, 2017} 

\title{Computational Social Choice and Computational Complexity: BFFs?\thanks{A version of this paper will appear in AAAI-2018~\cite{hem:ctoappear:bffs}.}}
\maketitle

\begin{abstract}
  We discuss the connection between computational social choice
  (comsoc) and computational complexity. We stress the work so far on,
  and urge 
continued 
focus on, two less-recognized aspects of this
  connection. Firstly, this is very much a two-way street: Everyone
  knows 
  complexity classification is used in comsoc, but we also 
  highlight benefits to complexity that have arisen from its use
  in comsoc. Secondly, more subtle, less-known complexity tools often
  can be very productively used in comsoc.
\end{abstract}

\section{Introduction}
Even in a work context, friendship is amazingly powerful.  We all know
how valuable it is to have collaborators who have each other's backs: working 
closely toward a shared project goal yet each bringing their own
skills and perspective to the project, 
to seek to reach 
insights and advances that neither 
researcher 
could have 
achieved
alone.

Research areas
can be like that too.  Computational social choice
(henceforward, comsoc)---which brings a 
computational-cost lens to such social-choice issues as 
preference aggregation/elections/manipulation/fair division---certainly has 
roots tracing far back; but comsoc's 
existence as a recognized,
distinctive research area within AI/multiagent-systems is quite
recent.  Despite that, its growth as an area during these recent years
has been explosive, and we 
suggest that a synergy between
comsoc and complexity 
has been partially responsible for that growth.
At the 2017 AAMAS conference~\cite{aamas17:c:aamas17}, for
example, there were four 
sessions devoted to Computational
Social Choice; no other topic had that many sessions.  
Those four sessions included 21 papers, and of those 21,
7 had the word ``complexity'' in their titles although 
in the entire conference only two other papers had the word complexity
in their titles.
In contrast, at
the 2003 AAMAS conference~\cite{aamas03:c:aamas03}, 
the string ``social choice''
does not even appear in the ACM Digital Library 
online table of contents;
neither 
does the string ``election'' or any form of the word ``vote,''
and only three papers in the entire conference have the word
``complexity'' in their titles.

One thing that those numbers are reflecting is the fact that 
comsoc and complexity 
have worked together very well.  
Comsoc
research, from the seminal work of Bartholdi, Orlin, Tovey, and Trick~\cite{bar-tov-tri:j:who-won,bar-tov-tri:j:manipulating,bar-orl:j:polsci:strategic-voting,bar-tov-tri:j:control}
onward, has done an enviably skilled job of interestingly employing
complexity classification---which admittedly has 
strengths and
weaknesses, both in its theory and in its application---to reveal a
subtle and varied landscape as to what is difficult and what is hard,
and how small changes in problems and models can induce seismic
changes in complexity.

Yet this article's primary goal 
is not 
to praise that praiseworthy
achievement.  It is to recognize two more subtle things that have
already happened, but that we think need to be recognized for what
they are, and what more can be done.

Firstly, the comsoc/computational complexity friendship is truly a
two-way street.  Although everyone knows that complexity
classification is widely used in comsoc research, we will stress a
direction that is not well-recognized within the AI community or the
complexity community: Research in comsoc has often been of great
benefit to complexity theory.  In particular, 
complexity classification 
within comsoc 
has populated key complexity
classes with problems that are undeniably \emph{natural} and
compelling.  Why is that important?  Complexity theorists like
creating complexity classes that capture computational
capabilities/resources that seem natural and compelling to the
complexity theorists.  But if those classes then turn out, for
example, to have no natural complete problems, it is like hosting a
party and having no one show up.  It could even make one 
wonder 
whether the classes really were important at all.  So complexity owes
quite a debt to comsoc, for bringing parties to life.  

In addition to stressing the two-way street aspect of the
comsoc/computational complexity friendship, we will also point out 
how rather subtle, lesser-known complexity
notions and machinery have in the comsoc world found fertile ground to
actually do important things.  
Complexity notions ranging from
search-versus-decision issues to one-to-one reductions to the join
operator have arguably 
been used in the comsoc world 
in
more natural and more satisfying 
ways than they have been used 
within complexity theory itself.

We 
present four different issues where the work in comsoc
exemplifies one or both of these themes: comsoc work helping
complexity, and lesser-known complexity notions and machinery being
used to address comsoc issues.  We will particularly stress examples
that fall in the area of the complexity of elections/voting, probably
the most intense current focus of comsoc and a remarkably challenging,
interesting, nuanced area (for a general survey of comsoc see the
excellent articles by 
Chevaleyre et al.~\cite{che-end-lan-mau:c:polsci-intro}
and 
Brandt, Conitzer, and Endriss~\cite{bra-con-end:b:comsoc},
and for surveys of elections/voting within comsoc
see the
surveys by Faliszewski et al.~\cite{fal-hem-hem-rot:b-abbrev:richer},
Faliszewski, Hemaspaandra, and 
Hemaspaandra~\cite{fal-hem-hem:j:cacm-survey},
and 
Faliszewski and Procaccia~\cite{fal-pro:j:war-on-manipulation}; 
and for additional
accessible surveys, see the relevant chapters in the important recent
textbook edited by Rothe~\cite{rot:b:econ}).
This article
gives 
just a few examples of these themes, but we try to discuss
each of those with some context and care, 
and in two places our 
examples will
connect complexity with other important AI areas,
planning and SAT solvers, once with the comsoc/complexity connection
growing from the other AI area and once with it creating 
an opportunity within
the other AI area.

Then in the conclusion we will suggest that the two 
themes we have been stressing and the
work to date on them
make clear that
comsoc experts and complexity experts 
working more closely and
extendedly together will be to the benefit of both fields.

\section{Make It a Party: Populating Lonely Complexity Classes}
As mentioned above, complexity classes are typically defined to
capture a certain mode/power/model of computing.  For example, the
important probabilistic complexity class 
BPP~\cite{gil:j:prob-tms}, 
bounded probabilistic
polynomial time,
is the class of those sets $A$ for which
there is a probabilistic polynomial-time Turing machine that on each input
agrees with $A$ with probability at least 75\% (or 51\%, or 
99\%;
all three of those turn out to define the same class).  The class
NP is the class of those sets accepted by nondeterministic
polynomial-time Turing machines.  The 
class UP~\cite{val:j:checking},
unambiguous polynomial time, 
is the
class of those sets $A$ accepted by nondeterministic polynomial-time
Turing machines whose acceptance is always unambiguous; for inputs not in $A$
the machine has no accepting computation paths, and for each input in
$A$, the machine has \emph{exactly one} accepting computation path (in
the lingo, ``has a unique succinct certificate'').

This section will present a few examples of classes that were not
wildly filled with natural sets that seem to 
capture their nature, 
yet
comsoc provided new, or even the first, such examples.  By ``capture
the nature'' of a class, in the dream case we would mean ``is complete
for the class,'' i.e., our set is in the class and has the property
that each set in the class many-one reduces to our set.
But if that can't be
achieved, finding a set that falls into a given class and does not
seem to fall into any smaller class is the natural 
fallback 
step to take.

Even readers not very familiar with complexity theory may know 
that there 
are thousands of known NP-complete problems and indeed a
whole book is devoted to such problems~\cite{gar-joh:b-special-abbreviated:int}; and that 
there is an entire book devoted even to problems complete
for P, with respect 
to the natural completeness type there
\cite{gre-hov-ruz:b:limits};
and that even for 
the levels of the polynomial hierarchy beyond $\np$ and $\conp$, such as 
$\np^\np$, $\conp^\np$,
$\np^{\np^\np}$, and $\conp^{\np^\np}$, there exists a 
detailed compendium~\cite{sch-uma:j:PH-part-one-with-web-updates-cited}.
And from that it would be easy to conclude that it is boring and easy 
to find natural problems complete for virtually any natural complexity class.

But that seems not to be so.  For example, 
$\np\cap \conp$
(the class of sets that have succinct, easily-checkable proofs both of
membership and nonmembership) and two of the classes mentioned above, BPP
and UP, are not known to have even one complete set---not a natural
one, and not even an unnatural
one~\cite{sip:c:complete-sets,har-hem:j:up}.  In fact, the just-cited
papers and the work of Regan~\shortcite{reg:unpub:cons} in effect are
showing that the existence of a complete set for one of these classes
would have sweeping consequences for our understanding of the class:
that the issue of whether each of these classes has a complete set is
in fact a disguised version of the question of whether there exists a
nice enumeration of machines that precisely 
covers the class, i.e.,
that provides what Regan calls a ``constructive programming system.''
But things are worse still.  It is now know that not only do these
classes seem to lack complete sets, but in fact, it is plausible
that---indeed there are black boxes (aka oracles) relative to
which---even far larger classes than the given class do not contain
even a single set that is many-one hard, or indeed even Turing hard,
for the class~\cite{hem-jai-ver:j:up-turing}.  

So it is not at all a
sure thing that important
classes---and BPP is undeniably important, since some would even argue
that BPP rather than P should be the working definition of
feasibility---have complete sets at all, much less natural complete
sets.

And yet some classes that had remained largely or wholly
unpopulated by natural sets needing the class's power 
\emph{have} had natural examples provided
by problems from comsoc, and other less empty but far from crowded
classes have also turned out to be what pinpoints the complexity of
important comsoc problems.

\subsection{$\boldthetatwo$: Parallel Access to NP}

Probably the most striking such example is the $\thetatwo$ level of
the polynomial hierarchy.  This class was introduced by Papadimitriou
and Zachos~\shortcite{pap-zac:c:two-remarks}, and is the class of sets
that can be accepted by polynomial-time machines allowed to make
$\bigo(\log n)$ sequential queries to an NP oracle; 
$\np \cup \conp \subseteq \thetatwo \subseteq \p^\np$.
Perhaps more
naturally, 
$\thetatwo$ is 
known to 
be the class of 
sets that can be
accepted by polynomial-time machines allowed to make (an unlimited
number of) nonadaptive (i.e., parallel) queries to an NP
oracle~\cite{hem:j:sky}.  For example, consider the language,
ParitySAT, that accepts its 
input exactly if the input is a list of Boolean formulas and the number 
of them that is satisfiable is itself an odd number.  
ParitySAT
is clearly in $\thetatwo$, since the polynomial-time machine can 
in parallel ask its oracle about the membership in SAT of each of its 
input formulas, and then can see whether the number that are satisfiable 
is odd. We will come back in a moment to the unnaturalness of 
this example and the search for natural examples.

$\thetatwo$ is a very important class in complexity theory.  
Briefly put, it has a large number of equivalent characterizations
that are quite natural, e.g., it is the class of 
sets that can be
accepted by logspace machines allowed to make an unlimited
number of adaptive queries to an NP
oracle~\cite{wag:j:bounded};
it is the class the polynomial hierarchy naturally is shown 
to collapse to if there are sparse (i.e., having at most 
a polynomial number of strings at each length) 
NP-Turing-complete sets~\cite{kad:j:pnplog};
and its relationship to $\p^\np$ is known
to completely characterize whether conversations with NP 
oracles can 
manufacture time-bounded randomness~\cite{hem-wec:j:man-rand}.

This at first sounds like quite a party, if one focuses on
the results $\thetatwo$ is central to.  But wait.  Even when all 
these results were known, the class was not known to have even one 
natural complete set.  It was known to have a large number of 
rather unnatural complete sets all having to do with counting the 
parity of items, e.g., the ParitySAT
example given above about telling whether 
out of a collection of Boolean formulas an odd number of them
are satisfiable is known to be complete for $\thetatwo$~%
\cite{wag:j:more-on-bh}.  

So in fact, as to known natural complete sets, the party was at that
time a (complexity-theoretically important but) empty room.  However,
comsoc put first one and then many people into that room.  In
particular, Charles Lutwidge Dodgson (aka Lewis Carroll) in the year
1876 had defined a 
fascinating 
election
system~\cite{dod:unpubMAYBE-without-embedded-citations:dodgson-voting-system}.
He was motivated by
the idea that it would be nice for an election system to ensure that
(a)~if (relative to the votes, which are assumed to each be a tie-free
total ordering of the candidates) there is a candidate---what is known
as a Condorcet winner~\cite{con:b:condorcet-paradox}---who is
preferred in head-on-head contests against each other candidate, then
that candidate is chosen the winner, and (b)~otherwise, each
candidate who is ``closest'' to being such a winner (in the sense
that the number of adjacent exchanges in preference orders needed to
make it become a Condorcet winner
is the lowest among the candidates) is named
a winner.  And Dodgson's system then does 
just that: It counts distance
from being a Condorcet winner, and the candidate(s) with the lowest
distance wins.  
In doing that, 
Dodgson was in the
1800s already using the notation that in modern computer science is central
and known as an edit distance.

Although Dodgson's system is mathematically very well defined, one of the
seminal papers of comsoc showed that the computational complexity of
implementing his system is not low.  In particular, Bartholdi, Tovey,
and Trick~\shortcite{bar-tov-tri:j:who-won} showed that it is NP-hard
to tell if a given candidate wins such an election.  They left open 
whether the problem is NP-complete, but eight years later the problem
was proven to be complete for 
$\thetatwo$~\cite{hem-hem-rot:j:dodgson}.  $\thetatwo$-complete
problems cannot be NP-complete unless the polynomial hierarchy 
collapses to $\np \cap \conp$, and thus 
the winner problem for Dodgson elections
is highly unlikely to be NP-complete.

But the most important thing to note here is that the complexity of
winner-testing for Dodgson elections is \emph{not} a problem that was
rigged to provide a $\thetatwo$-complete set.  The election system was
defined in the 1800s for its own interest and importance, long before
NP was ever dreamed of, much less $\thetatwo$.  
This problem---from 
comsoc---thus
provides an extremely natural $\thetatwo$-complete set.

This result sparked interest in whether other problems in comsoc and
beyond might also be $\thetatwo$-complete.  And the floodgates opened.
Other important election systems such as those of Kemeny and Young
were proven to also be
$\thetatwo$-complete~\cite{hem-spa-vog:j:kemeny,rot-spa-vog:j:young}.
General tools were extracted from this approach to try to make it
easier to prove such results~\cite{spa-vog:c:theta-two-classic},
$\thetatwo$-completeness was shown to capture the complexity of how
well certain greedy algorithms 
do~\cite{hem-rot:j:max-independent-set-by-greedy},
and a survey was written looking at the meaning of improving from
NP-hardness results to $\thetatwo$-completeness
results~\cite{hem-hem-rot:j:raising-lower-bounds}.  Even in the quite
different field of automata theory, $\thetatwo$-complete problems were
found to capture important, natural
notions~\cite{hol-mac:c:empty-alternating-stack-automata}.  Briefly
put, the party was 
very much on!

\subsection{\boldmath $\np^\pp$ and Other Classes}
There are many other classes that comsoc and other work within AI have
helped populate.  One of the most striking examples is the class
$\np^\pp$, i.e., the sets solvable by nondeterministic polynomial-time
Turing machines given access to (their choice of a) set from PP
(probabilistic polynomial time, the class of sets for which there is a
probabilistic polynomial-time Turing machine that on each input is
correct with probability greater than 50\%).  $\np^\pp$ is a class of
great
complexity, and it is not always easy to
work with.  In terms of descriptive generality and its ``location,''
Toda's~\shortcite{tod:j:pp-ph} 
Theorem says that $\p^\pp$
contains not just the polynomial hierarchy ($\ph$) but even
$\pp^\ph$.  So $\np^\pp$ contains both of those and even
$\np^{\pp^\ph}$, while 
being itself contained in $\pspace$.

It was only in the late 1990s that people started finding 
natural complete
sets for $\np^\pp$, and that work came not from 
complexity issues but from an AI domain, 
the study of planning and
decision processes.  There are a number of papers on this 
and we won't
here cite them all, but we mention as a pointer 
the very impressive work of
Goldsmith, Littman, and Mundhenk~\shortcite{gol-lit-mun:j:planning},
which shows versions of both the plan-existence problem and the
plan-evaluation problem are, for partially ordered plans,
$\np^\pp$-complete, and 
of Mundhenk et
al.~\shortcite{all-gol-lus-mun:j:hidden-markov} on Markov decision
processes, which also yields natural $\np^\pp$-complete problems.

Comsoc also has helped populate that party.  Although currently its
$\np^\pp$ results are upper bounds---and we commend to the reader the
open issue of either proving completeness or of lowering the upper
bound---the beautiful work of Mattei, Goldsmith, Klapper, and
Mundhenk~\shortcite{gol-mat-mun-kla:j:bribery-tournaments-uncertain} shows that
$\np^\pp$ plays an important role in the study of bribery and manipulation
in tournaments of uncertain information, by showing that many such
problems are in $\np^\pp$.\footnote{After doing that, that paper comments
that ``we have actually shown these problems are in $\np^{\pp[1]}$,''
which is 
the class in which the NP machine is allowed at most one query per path
to the PP oracle.  Simply so that comment in the paper doesn't
lead any reader of the present paper to 
think that
$\np^\pp$-completeness for that paper's problems is unlikely due to
that paper having achieved the seemingly better upper bound of
$\np^{\pp[1]}$, we claim that in fact
$\np^{\pp[1]} = \np^{\pp}$.  Let us prove that claim.
  $\np^{\pp[1]} = \np^{\pp}$ holds via altering the NP machine of
  $\np^\pp$ to instead guess the answer to each PP query (without
  asking the query) and then via a single truth-table (parallel) query
  round to PP checking that all its guesses were correct, and if a
  given path would have accepted if its guesses of the answers were
  correct and that path finds via its truth-table-round access to PP
  that it guessed correctly, then that path accepts.  That as just stated
  is illegally overly using the PP oracle. But since
  PP is closed
  under truth-table reductions~\cite{for-rei:j:pp}, the truth-table round can be replaced
  with a single query, and thus we indeed have establishing the
  claimed class equality. Indeed, since the truth-table yields True
  for only 
  one known-at-truth-table-query-time setting of answers, even
  Beigel, Reingold, and Spielman~\shortcite[p.~195]{bei-rei-spi:j:pp-closure}
  would suffice to allow a single query per path.}

There are many other parties, some admittedly not so previously poorly
attended as those thrown by $\thetatwo$ and $\np^\pp$, where comsoc
has provided attendees.  Here are just a few examples.  Nguyen et
al.~\shortcite{ngu-ngu-roo-rot:j:dp-multiagent-resource-allocation}
gave the class DP (all sets that are the symmetric difference of two
NP sets) a wide range of new complete sets having to do with social
welfare optimization in multiagent resource allocation.  The first
$\p^{\np[1]}$-completeness result in social choice (which also is perhaps
the first natural 
completeness result anywhere for 
$\p^{\np[1]}$) and the first $\p^{\np}$-completeness result in social
choice both come from a study of the complexity of weighted
coalitional manipulation of elections, under the voting system, Veto,
where each voter vetos one candidate and whoever has the fewest vetos
wins; in an online setting (what in political science is called a
roll-call vote), the 
winner problem here is
$\p^{\np[1]}$-complete for 3 candidates and is $\p^{\np}$-complete
for 4 or more candidates~\cite{hem-hem-rot:j:online-manipulation}.  
Filos-Ratsikas and Goldberg~\shortcite{fil-gol:t:consensus}
have recently proven 
a natural cake-cutting-related problem to be complete for 
the lonely class PPA\@.
And a huge
number of papers in comsoc contribute natural problems---often related
to manipulative attacks on elections---to other classes such as $\np^\np$,
$\conp^\np$, and PSPACE\@.  
Also, many counting-based
issues in comsoc have been shown to be complete for the counting
analogue of NP, 
the class $\sharpp$.  

In summary, there is a two-way street of friendship between comsoc and
computational complexity.  Complexity benefits since its classes---some of
which played central roles in key abstract results yet had few or
no natural complete problems---%
are given 
quite
compelling 
natural problems based in comsoc.  Comsoc in turn
has benefited 
since the machinery of complexity-class classification 
helps
clarify how hard or easy many of its problems are.

\section{Complexity Machinery and Notions 
Find 
Fertile Ground 
in Comsoc}

This section provides a 
tour of 
two cases where
complexity techniques and notions have found very fertile ground in
comsoc research.  
Although 
one-to-one 
(Section~\ref{one-to-one}) reductions have been 
used centrally in complexity theory, 
in particular in the 1970s, 
their new use in comsoc is both powerful and 
quite different than the use they had in complexity.
And in another case
(Section~\ref{s-vs-d}), 
comsoc has given new life to lovely complexity
  machinery that had been developed in 1976, and yet that even 36
  years later had not found a single application in a
real-world 
domain.

We will cover each topic but briefly,
  though we will try to give a sense of how each supports this
  section's thesis that comsoc has been a fertile ground for complexity
  techniques and notions, helping comsoc of course, but also in the
  case of little-used or narrowly used complexity techniques helping establish
  the value or breadth of the techniques.

\subsection{Search versus Decision for NP Problems}\label{s-vs-d}
Everyone knows that SAT is a decision problem, and indeed that the
vast majority of complexity classes are defined in terms of decision
problems, not search problems.  This has been the case for so long
that it is easy to forget 
\emph{why} this is so.  

The reason why complexity is largely focused on decision problems is
because for almost all natural problems the search problem clearly
polynomial-time Turing reduces (and often, for example for SAT, even
``polynomial-time 2-disjunctive truth-table'' reduces) to the problem's
decision version.  And so the two problems are inherently of the same
complexity, give or take composition with a polynomial; 
it is impossible in these cases
for decision to be easy yet for search to be hard.  SAT is the classic
case: Given a Boolean formula, $F(v_1,v_2,\ldots)$, it is satisfiable
exactly if at least one of $F(\mbox{\sc True},v_2,\ldots)$ and $F(\mbox{\sc False},v_2,\ldots)$ is
satisfiable, so given a decision procedure for SAT, one can ask those
two questions to it, and thus find a good value for the first variable (if
any exists), and then can redo the process, on the ``reduced''
formula, to similarly set the second variable and so on.  By the end,
using as a black-box a decision procedure for SAT, one has found a
satisfying assignment in polynomial time, when one exists.

Are all natural NP problems so polite as to allow themselves to be
fitted into this straitjacket of linked complexities?  For
decades there have been hints that the answer might be ``no.''  Most
powerfully, in 1976 Borodin and
Demers~\shortcite{bor-dem:t-abbrev:selfreducibility} built lovely,
clever machinery that showed that if
$\p \neq \np \cap \conp $, then there exists an infinite 
polynomial-time recognizable
subset of SAT
that has no 
polynomial-time search algorithm finding solutions for its members.  
In brief, one could quickly determine with certainty that its 
members were satisfiable, but one has no idea---it turns out not even 
a good setting for the first variable of the formula---of what 
a satisfying assignment would be.
(The reason this claim is not
precluded by the above discussion of self-reducibility is that the 
Borodin--Demers 
subset is so pathological that it has infinitely many members such
that the formulas generated at some stage of the above self-reduction
process are all not in the subset, and so the above process is
rendered invalid on the subset.)

Stunning though that result was, the fact that it was achieving its
power not on all of SAT (which is impossible), but rather on a
somewhat pathological subset of SAT, 
kept the result and
its lovely machinery from gaining traction; it remained only as the
above 1976 technical report, and did not have a conference or a
journal version.  The work was likely---and in the first 
and second of these four examples the work is explicitly discussed---the 
inspiration behind some
later related work, e.g., that for unambiguous computation the 
implication mentioned above becomes an ``if and only if'' 
characterization~\cite{har-hem:j:up},
that under the extremely strong assumption 
that deterministic and nondeterministic {double exponential time}
differ there exists an (artificial) language in NP for which 
search does not reduce to decision~\cite{bel-gol:j:s-vs-d},
that if $\p \neq \np$ then there exist NP-complete languages 
that are not self-reducible~\cite{fal-ogi:j:self},
and
that relative to some oracle there is a search versus decision separation
for exponential-time classes~\cite{imp-tar:c:dec}.

Nonetheless, the work found not a single application on natural
problems for 36 years.  And it was comsoc that finally provided that
application.  In particular, Hemaspaandra, Hemaspaandra, and
Menton~\shortcite{hem-hem-men:c-abbreviated:search-versus-decision} proved that of
the standard types of attacks on elections, for about half it holds
that if $\p \neq \np \cap \conp$ then there is an election system
having a polynomial-time winner problem for which the decision problem
for that type of attack is in P but the search problem cannot be
solved in polynomial time.  (For the other half of the attack
types, that paper proved unconditionally that search \emph{does}
reduce to decision.)  Note that the attack types were not created by
that paper; they are the long-standard attack types.  And so the
problems themselves are not tricks, but are the natural, standard
ones, and on these, search and decision are being separated under the
given complexity-theoretic assumption, which is widely believed to be
true.

There are four comments which pretty compellingly need to be 
made 
at this point.  
First, if $\p = \np \cap \conp$, then integer
factoring can be done in polynomial time, and so RSA and much of the
foundation of modern computer security falls.  So the above results
say that either the foundations of cryptography are deeply flawed, or
the approach taken within comsoc to defining election-attack
problems---namely, they are defined in their decision versions---is
defining as tractable some problems whose search 
version 
(the issue of
finding the attack that works) is not tractable.  Second, the reason
this separation of search and decision is possible is that, quite
unexpectedly, about half of the natural election-attack problems do
not seem to be nicely self-reducible.
Third, natural is in the eye of the
beholder; although the election-attack problems used in the result are
the standard, natural ones, the election systems used are indirectly
specified by the hypothesized sets in $(\np \cap \conp) - \p$, and so
are likely not very natural.  And fourth and finally, this entire approach
then found application in a different area of AI, namely, the study
of backbones and backdoors of Boolean 
formulas, where it was shown for 
example that if 
$\p \neq \np \cap \conp $, then search versus decision separations 
occur for backbones~\cite{hem-nar:c:backbones-opacity,hem-nar:t-abbrev:backdoors-opacity}.

\subsection{Density-of-Hardness Transfer and 
One-to-One Reductions}\label{one-to-one}
When one says a problem is NP-complete, that doesn't prove that it is
hard.  It just proves that if anything in NP is hard, then that problem is
hard.  This is generally viewed as much better than having nothing to say
about the difficulty of the problem.

Wouldn't it be 
lovely to have an analogous type of result for frequency
of hardness?  For example, in Section~\ref{s-vs-d}, we discussed work
showing that if $\p \neq \np \cap \conp $, then search 
and 
decision separate for many key issues regarding manipulative attacks
on elections and backbones of Boolean formulas.  But maybe in those examples,
for which we know decision is easy, the hardness of search may be a 
con: Perhaps search is only hard on a very, very, very small portion of
the inputs, asymptotically?  That is, perhaps it is a worst-case separation
that doesn't hold in the typical case.

To try to argue against that possibility, it would be wonderful if we
could argue that if \emph{even one} problem $A$ in $\np \cap \conp$
is hard with a certain frequency $h$ (i.e., each polynomial-time
heuristic is wrong on $A$'s membership problem 
on $\Omega(h(n))$
strings up to length $n$) then every polynomial-time
heuristic for our search problem fails with almost that same
frequency, namely, for each polynomial-time heuristic there is an
$\epsilon > 0$ such that on our given search problem it fails on 
$\Omega(h(n^\epsilon))$ of the strings up to length $n$).

This would say that if \emph{any} set in $\p \neq \np \cap \conp$ is
frequently hard (relative to heuristic attacks), then our search
versus decision separations are not cons.  After all, it is widely
believed, most crucially in the cryptography community (due in part to
the issue of factoring), that there are sets in $\np\cap\conp$ that
\emph{are} frequently hard with respect to polynomial-time heuristics.  And
if one believes that, then the hypothetical machinery discussed above
would 
immediately 
convert that into a claim of the frequent hardness
of the created search problems whose decision problems are easy.

In fact, machinery that 
can 
do this exists, although it was developed for 
a completely different purpose in the 1970s.  So this is an example where
comsoc---and the study of SAT solvers---is giving a fresh use to 
a complexity-theoretic notion.

In particular, every CS undergraduate learns why claims of
NP-completeness are crucially tied to the tool of many-one
polynomial-time reductions; it is how we establish them.  In the
1970s, though, there was a sudden moment of focus in theoretical
computer science on the notion of \emph{one-to-one} (aka {injective})
polynomial-time reductions, i.e., polynomial-time reductions that have
no collisions.  

This moment of intense focus on one-to-one reductions came about to support
one of the great complexity goals of the 1970s: to prove that all
NP-complete sets were the same set in a transparent disguise, namely,
that all NP-complete sets were polynomial-time isomorphic.  This is
known as the 
Isomorphism Conjecture (aka~the Berman--Hartmanis Isomorphism Conjecture).  
To this day it
remains open.
However, using one-to-one reductions,
in the 1970s Berman and Hartmanis~\shortcite{ber-har:j:iso} proved
that all familiar NP-complete sets indeed were polynomial-time
isomorphic, which was  
a huge
revelation.  
(Their insight
in using one-to-one reductions 
was motivated by the fact that in recursive
function theory, recursive one-to-one reductions are central to the
proof that all RE-complete sets are recursively isomorphic, a result
that follows from the 
so-called Myhill~\shortcite{myh:j:creative} Isomorphism Theorem.)
However, due to challenges such as 
the potential existence of one-way functions, that 1970s result has 
not only never been expanded to (under the inherently required assumption
that $\p \neq \np$) include all NP-complete sets, but indeed there is 
evidence that there may well exist nonisomorphic NP-complete
sets~\cite{jos-you:j:kcreative,kur-mah-roy:j:prob1}, i.e., that the 
Isomorphism Conjecture may fail.

These days in complexity theory, many-one reductions remain the
standard, and one-to-one reductions are not often discussed.  However,
one thing that one-to-one reductions do fiendishly well is preserve
density.  If one has a bunch of strings with a certain behavior, their
image when pushed through the reduction cannot possibly jumble them on
top of each other, because one-to-one reductions don't jumble anything
on top of anything else.  It is admittedly true that such reductions
can leave ``holes'' (though injective, they need not be surjective),
and can polynomially stretch (and superpolynomially contract, though
inherently not too frequently) lengths; and that in fact is what is
behind the ``$\epsilon$'' mentioned above.  But one-to-one 
reductions by definition
never have collisions.

Exploiting that, the ``if anything in $\np \cap \conp$ is frequently hard
relative to polynomial-time heuristics then our search problems 
are almost as
frequently hard relative to polynomial-time heuristics as those''
results mentioned above
are achieved by 
ensuring that the entire proof
structure in those papers can be made to yield polynomial-time
one-to-one reductions.  (And the additional related paper mentioned 
earlier, which adds in
results on the case of back\emph{doors} of Boolean formulas, and has
some related results about backbones of Boolean formulas, in both
cases under the weaker assumption $\p \neq \np$, also works by
achieving polynomial-time one-to-one reductions.)

Thus, to summarize, one-to-one reductions 
are a tool that was briefly extremely important in
complexity in the 1970s in a brilliant but even forty years 
later 
not yet successful attempt to show that there is in effect just
one NP-complete set.  But comsoc---and also, related to SAT solvers,
the study of backbones and backdoors of Boolean formulas---has
brought the notion of one-to-one reductions 
again front-and-center,
providing an ``if anything in the class is frequently hard then
\emph{this} is frequently hard'' analogue of the ``if anything in the
class is hard then \emph{this} is hard'' argument that many-reductions
have provided for almost half a century in the case of NP-complete
sets.  

This case makes very clear 
the two-way street that
exists between complexity and comsoc (and also the study of SAT
solvers).  Complexity benefits in that one of its notions that was a
bit covered in cobwebs turned out to yield important
frequency-of-hardness results in comsoc, and also relatedly in the study
of SAT solvers.  And comsoc benefits in that it now has these results,
which are quite powerful evidence that the hardness of search that
these speak of is in fact not some con job 
that
happens only extremely infrequently.

\section{Conclusions}
We have discussed just a few of the rich collection of interactions
between comsoc and computational complexity.  
Many others---from 
how the complexity-theoretic join
operation allows proofs in comsoc of the impossibility of proving
certain impossibility 
theorems~\cite{hem-hem-rot:j:hybrid} to 
how work on online control gives an
unexpected new quantifier-alternation 
characterization of the class 
coNP~\cite{hem-hem-rot:j:online-voter-control} to 
the rich interactions with approximation (see as just one of 
many examples 
\cite{fal-sko-tal:c-abbrev:bribery-multiwinner})
to
the power of dichotomy results to the insights given 
by parameterized complexity (both these last are on view 
simultaneously, for example, in 
\cite{dey-mis-nar:c-abbrev:parameterized-dichotomy-bribery})
to 
much 
more---are not even
touched on here.  

What are the take-aways?  The benefits of the friendship between
comsoc and computational complexity are a two-way street; both comsoc
and complexity have benefited.  Classes from complexity have been
populated, cobweb-covered techniques have been given surprising new
life, and interesting results have been obtained in comsoc from the
use of those classes and techniques.  In light of this, we urge
researchers in comsoc and complexity to reach out and work with each
other more and more, for the benefit of both fields.  
And we hope that more Ph.D. programs will encourage and create
young researchers who are trained in and simultaneously expert in
\emph{both} areas; that will allow advances far beyond even those that
so far have come from this wonderful synergy
between areas.  

Let us hope that the areas 
become best friends forever.

\subsubsection*{Acknowledgments} 
Warm thanks to 
the anonymous AAAI-18 reviewers, 
Aris Filos-Ratsikas, and Paul Goldberg for helpful 
comments.

\bibliographystyle{alpha}
%
%

%
%
%
%
%
%
%


\begin{thebibliography}{MGKM15}

\bibitem[BCE13]{bra-con-end:b:comsoc}
F.~Brandt, V.~Conitzer, and U.~Endriss.
\newblock Computational social choice.
\newblock In G.~Weiss, editor, {\em Multiagent Systems}, pages 213--284. MIT
  Press, 2nd edition, 2013.

\bibitem[BD76]{bor-dem:t-abbrev:selfreducibility}
A.~Borodin and A.~Demers.
\newblock Some com\-ments on func\-tional self-re\-du\-ci\-bil\-ity and the
  {N}{P} hier\-ar\-chy.
\newblock Technical Report TR 76-284, Dept.\ of Computer Science, Cornell
  University, July 1976.

\bibitem[BG94]{bel-gol:j:s-vs-d}
M.~Bellare and S.~Goldwasser.
\newblock The complexity of decision versus search.
\newblock {\em SIAM Journal on Computing}, 23(1):97--119, 1994.

\bibitem[BH77]{ber-har:j:iso}
L.~Berman and J.~Hartmanis.
\newblock On isomorphisms and density of {{N}{P}} and other complete sets.
\newblock {\em SIAM Journal on Computing}, 6(2):305--322, 1977.

\bibitem[BO91]{bar-orl:j:polsci:strategic-voting}
J.~{{Bartholdi}}, III and J.~Orlin.
\newblock Single transferable vote resists strategic voting.
\newblock {\em Social Choice and Welfare}, 8(4):341--354, 1991.

\bibitem[BRS95]{bei-rei-spi:j:pp-closure}
R.~Beigel, N.~Reingold, and D.~Spielman.
\newblock {PP} is closed under intersection.
\newblock {\em Journal of Computer and System Sciences}, 50(2):191--202, 1995.

\bibitem[BTT89a]{bar-tov-tri:j:manipulating}
J.~{{Bartholdi}}, III, C.~Tovey, and M.~Trick.
\newblock The computational difficulty of manipulating an election.
\newblock {\em Social Choice and Welfare}, 6(3):227--241, 1989.

\bibitem[BTT89b]{bar-tov-tri:j:who-won}
J.~{{Bartholdi}}, III, C.~Tovey, and M.~Trick.
\newblock Voting schemes for which it can be difficult to tell who won the
  election.
\newblock {\em Social Choice and Welfare}, 6(2):157--165, 1989.

\bibitem[BTT92]{bar-tov-tri:j:control}
J.~{{Bartholdi}}, III, C.~Tovey, and M.~Trick.
\newblock How hard is it to control an election?
\newblock {\em Mathematical and Computer Modeling}, 16(8/9):27--40, 1992.

\bibitem[CELM07]{che-end-lan-mau:c:polsci-intro}
Y.~Chevaleyre, U.~Endriss, J.~Lang, and N.~Maudet.
\newblock A short introduction to computational social choice.
\newblock In {\em Proceedings of the 33rd International Conference on Current
  Trends in Theory and Practice of Computer Science}, pages 51--69.
  Springer-Verlag {\it Lecture Notes in Computer Science \#4362}, January 2007.

\bibitem[Con85]{con:b:condorcet-paradox}
{J.-A.-N. de Caritat, Marquis de} Condorcet.
\newblock {\em Essai sur l'Application de L'Analyse \`{a} la Probabilit\'{e}
  des D\'{e}cisions Rendues \`{a} la Pluralit\'{e} des Voix}.
\newblock 1785.
\newblock Facsimile reprint of original published in Paris, 1972, by the
  Imprimerie Royale.

\bibitem[DMN17]{dey-mis-nar:c-abbrev:parameterized-dichotomy-bribery}
P.~Dey, N.~Misra, and Y.~Narahari.
\newblock Parameterized dichotomy of choosing committees based on approval
  votes in the presence of outliers.
\newblock In {\em Proceedings of the 16th International Conference on
  Autonomous Agents and Multiagent Systems}, pages 42--50, May 2017.

\bibitem[Dod76]{dod:unpubMAYBE-without-embedded-citations:dodgson-voting-system}
C.~Dodgson.
\newblock A method of taking votes on more than two issues.
\newblock Pamphlet printed by the Clarendon Press, Oxford, and headed ``not yet
  published'', 1876.

\bibitem[FG17]{fil-gol:t:consensus}
A.~{Filos-Ratsikas} and P.~Goldberg.
\newblock {Consensus Halving} is {PPA}-complete.
\newblock Technical Report arXiv:1711.04503~[cs.CC], Computing Research
  Repository, \mbox{arXiv.org/corr/}, November 2017.

\bibitem[FHH10]{fal-hem-hem:j:cacm-survey}
P.~Faliszewski, E.~Hemaspaandra, and L.~Hemaspaandra.
\newblock Using complexity to protect elections.
\newblock {\em Communications of the ACM}, 53(11):74--82, 2010.

\bibitem[FHHR09]{fal-hem-hem-rot:b-abbrev:richer}
P.~Faliszewski, E.~Hemaspaandra, L.~Hemaspaandra, and J.~Rothe.
\newblock A richer understanding of the complexity of election systems.
\newblock In S.~Ravi and S.~Shukla, editors, {\em Fundamental Problems in
  Computing}, pages 375--406. Springer, 2009.

\bibitem[FO10]{fal-ogi:j:self}
P.~Faliszewski and M.~Ogihara.
\newblock On the autoreducibility of functions.
\newblock {\em Theory of Computing Systems}, 46(2):222--245, 2010.

\bibitem[FP10]{fal-pro:j:war-on-manipulation}
P.~Faliszewski and A.~Procaccia.
\newblock {AI's} war on manipulation: {Are} we winning?
\newblock {\em AI Magazine}, 31(4):53--64, 2010.

\bibitem[FR96]{for-rei:j:pp}
L.~Fortnow and N.~Reingold.
\newblock {PP} is closed under truth-table reductions.
\newblock {\em Information and Computation}, 124(1):1--6, 1996.

\bibitem[FST17]{fal-sko-tal:c-abbrev:bribery-multiwinner}
P.~Faliszewski, P.~Skowron, and N.~Talmon.
\newblock Bribery as a measure of candidate success: {Complexity} results for
  approval-based multiwinner rules.
\newblock In {\em Proceedings of the 16th International Conference on
  Autonomous Agents and Multiagent Systems}, pages 6--14, May 2017.

\bibitem[GHR95]{gre-hov-ruz:b:limits}
R.~Greenlaw, H.~Hoover, and W.~Ruzzo.
\newblock {\em Limits to Parallel Computation: {P}-Completeness Theory}.
\newblock Oxford University Press, 1995.

\bibitem[Gil77]{gil:j:prob-tms}
J.~Gill.
\newblock Computational complexity of probabilistic {Turing} machines.
\newblock {\em SIAM Journal on Computing}, 6(4):675--695, 1977.

\bibitem[GJ79]{gar-joh:b-special-abbreviated:int}
M.~Garey and D.~Johnson.
\newblock {\em Computers and Intractability: {A} Guide to the Theory of
  {NP}-Completeness}.
\newblock {W. H. Freeman}, 1979.

\bibitem[Hem89]{hem:j:sky}
L.~Hemachandra.
\newblock The strong exponential hierarchy collapses.
\newblock {\em Journal of Computer and System Sciences}, 39(3):299--322, 1989.

\bibitem[Hem18]{hem:ctoappear:bffs}
L.~Hemaspaandra.
\newblock Computational social choice and computational complexity: {BFFs}?
\newblock In {\em Proceedings of the 32nd AAAI Conference on Artificial
  Intelligence}. AAAI Press\typeout{Minor Panic: missing pages AND volume if
  any, and REMOVE the to-appear note}, February 2018.
\newblock To appear.

\bibitem[HH88]{har-hem:j:up}
J.~Hartmanis and L.~Hemachandra.
\newblock Complexity classes without machines: {O}n complete languages for
  {UP}.
\newblock {\em Theoretical Computer Science}, 58(1--3):129--142, 1988.

\bibitem[HHM13]{hem-hem-men:c-abbreviated:search-versus-decision}
E.~Hemaspaandra, L.~Hemaspaandra, and C.~Menton.
\newblock Search versus decision for election manipulation problems.
\newblock In {\em Proceedings of the 30th Annual Symposium on Theoretical
  Aspects of Computer Science}, pages 377--388, 2013.

\bibitem[HHR97a]{hem-hem-rot:j:dodgson}
E.~Hemaspaandra, L.~Hemaspaandra, and J.~Rothe.
\newblock Exact analysis of {D}odgson elections: {L}ewis {C}arroll's 1876
  voting system is complete for parallel access to {NP}.
\newblock {\em Journal of the ACM}, 44(6):806--825, 1997.

\bibitem[HHR97b]{hem-hem-rot:j:raising-lower-bounds}
E.~Hemaspaandra, L.~Hemaspaandra, and J.~Rothe.
\newblock Raising {NP} lower bounds to parallel {NP} lower bounds.
\newblock {\em SIGACT News}, 28(2):2--13, 1997.

\bibitem[HHR09]{hem-hem-rot:j:hybrid}
E.~Hemaspaandra, L.~Hemaspaandra, and J.~Rothe.
\newblock Hybrid elections broaden complexity-theoretic resistance to control.
\newblock {\em Mathematical Logic Quarterly}, 55(4):397--424, 2009.

\bibitem[HHR14]{hem-hem-rot:j:online-manipulation}
E.~Hemaspaandra, L.~Hemaspaandra, and J.~Rothe.
\newblock The complexity of online manipulation of sequential elections.
\newblock {\em Journal of Computer and System Sciences}, 80(4):697--710, 2014.

\bibitem[HHR17]{hem-hem-rot:j:online-voter-control}
E.~Hemaspaandra, L.~Hemaspaandra, and J.~Rothe.
\newblock Online voter control in sequential elections.
\newblock {\em Autonomous Agents and Multi-Agent Systems}, 31(5):1055--1076,
  May 2017.

\bibitem[HJV93]{hem-jai-ver:j:up-turing}
L.~Hemaspaandra, S.~Jain, and N.~Vereshchagin.
\newblock Banishing robust {T}uring completeness.
\newblock {\em International Journal of Foundations of Computer Science},
  4(3):245--265, 1993.

\bibitem[HM00]{hol-mac:c:empty-alternating-stack-automata}
M.~Holzer and P.~McKenzie.
\newblock Alternating and empty alternating auxiliary stack automata.
\newblock In {\em Proceedings of the 25th International Symposium on
  Mathematical Foundations of Computer Science}, pages 415--425.
  Springer-Verlag {\it Lecture Notes in Computer Science \#1893},
  August/September 2000.

\bibitem[HN17a]{hem-nar:c:backbones-opacity}
L.~Hemaspaandra and D.~Narv\'{a}ez.
\newblock The opacity of backbones.
\newblock In {\em Proceedings of the 31st AAAI Conference on Artificial
  Intelligence}, pages 3900--3906. AAAI Press, February 2017.

\bibitem[HN17b]{hem-nar:t-abbrev:backdoors-opacity}
L.~Hemaspaandra and D.~Narv\'{a}ez.
\newblock The opacity of backbones and backdoors under a weak assumption.
\newblock Technical Report arXiv:1706.04582~[cs.AI], Computing Research
  Repository, \mbox{arXiv.org/corr/}, June 2017.

\bibitem[HR98]{hem-rot:j:max-independent-set-by-greedy}
E.~Hemaspaandra and J.~Rothe.
\newblock Recognizing when greed can approximate maximum independent sets is
  complete for parallel access to {N}{P}.
\newblock {\em Information Processing Letters}, 65(3):151--156, 1998.

\bibitem[HSV05]{hem-spa-vog:j:kemeny}
E.~Hemaspaandra, H.~Spakowski, and J.~Vogel.
\newblock The complexity of {Kemeny} elections.
\newblock {\em Theoretical Computer Science}, 349(3):382--391, 2005.

\bibitem[HW91]{hem-wec:j:man-rand}
L.~Hemachandra and G.~Wechsung.
\newblock {K}olmogorov characterizations of complexity classes.
\newblock {\em Theoretical Computer Science}, 83:313--322, 1991.

\bibitem[IT89]{imp-tar:c:dec}
R.~Impagliazzo and G.~Tardos.
\newblock Decision versus search problems in super-polynomial time.
\newblock In {\em Proceedings of the 30th IEEE Symposium on Foundations of
  Computer Science}, pages 222--227. IEEE Computer Society Press,
  October/November 1989.

\bibitem[JY85]{jos-you:j:kcreative}
D.~Joseph and P.~Young.
\newblock Some remarks on witness functions for non-polynomial and non-complete
  sets in {N}{P}.
\newblock {\em Theoretical Computer Science}, 39(2--3):225--237, 1985.

\bibitem[Kad89]{kad:j:pnplog}
J.~Kadin.
\newblock $\rm {P}^{{NP}[\log {\it n}]}$ and sparse {T}uring-complete sets for
  {N}{P}.
\newblock {\em Journal of Computer and System Sciences}, 39(3):282--298, 1989.

\bibitem[KMR95]{kur-mah-roy:j:prob1}
S.~Kurtz, S.~Mahaney, and J.~Royer.
\newblock The {Isomorphism} {Conjecture} fails relative to a random oracle.
\newblock {\em Journal of the ACM}, 42(2):401--420, 1995.

\bibitem[LGM98]{gol-lit-mun:j:planning}
M.~Littman, J.~Goldsmith, and M.~Mundhenk.
\newblock The computational complexity of probabilistic planning.
\newblock {\em Journal of Artificial Intelligence Research}, 9:1--36, 1998.

\bibitem[LWDD17]{aamas17:c:aamas17}
K.~Larson, M.~Winikoff, S.~Das, and E.~Durfee, editors.
\newblock {\em Proceedings of the 16th International Conference on Autonomous
  Agents and Multiagent Systems, {AAMAS} 2017, S{\~{a}}o Paulo, Brazil, May
  8--12, 2017}. International Foundation for Autonomous Agents and Multiagent
  Systems, 2017.

\bibitem[MGKM15]{gol-mat-mun-kla:j:bribery-tournaments-uncertain}
N.~Mattei, J.~Goldsmith, A.~Klapper, and M.~Mundhenk.
\newblock On the complexity of bribery and manipulation in tournaments with
  uncertain information.
\newblock {\em Journal of Applied Logic}, 13(4, Part 2):557--581, 2015.

\bibitem[MGLA00]{all-gol-lus-mun:j:hidden-markov}
M.~Mundhenk, J.~Goldsmith, C.~Lusena, and E.~Allender.
\newblock Complexity of finite-horizon {Markov} decision process problems.
\newblock {\em Journal of the ACM}, 47(4):681--720, July 2000.

\bibitem[Myh55]{myh:j:creative}
J.~Myhill.
\newblock Creative sets.
\newblock {\em Zeitschrift {f{\"u}r} Mathematische Logik und Grundlagen der
  Mathematik}, 1:97--108, 1955.

\bibitem[NNRR14]{ngu-ngu-roo-rot:j:dp-multiagent-resource-allocation}
N.~Nguyen, T.~Nguyen, M.~Roos, and J.~Rothe.
\newblock Computational complexity and approximability of social welfare
  optimization in multiagent resource allocation.
\newblock {\em Autonomous Agents and Multi-Agent Systems}, 28(2):256--289,
  2014.

\bibitem[PZ83]{pap-zac:c:two-remarks}
C.~Papadimitriou and S.~Zachos.
\newblock Two remarks on the power of counting.
\newblock In {\em Proceedings of the 6th GI Conference on Theoretical Computer
  Science}, pages 269--276. Springer-Verlag {\it Lecture Notes in Computer
  Science \#145}, January 1983.

\bibitem[Reg89]{reg:unpub:cons}
K.~Regan.
\newblock Provable complexity properties and constructive reasoning.
\newblock Manuscript, April 1989.

\bibitem[Rot16]{rot:b:econ}
J.~Rothe, editor.
\newblock {\em Economics and Computation: {An} Introduction to Algorithmic Game
  Theory, Computational Social Choice, and Fair Division}.
\newblock Springer, 2016.

\bibitem[RSV03]{rot-spa-vog:j:young}
J.~Rothe, H.~Spakowski, and J.~Vogel.
\newblock Exact complexity of the winner problem for {Young} elections.
\newblock {\em Theory of Computing Systems}, 36(4):375--386, 2003.

\bibitem[RWSY03]{aamas03:c:aamas03}
J.~Rosenschein, M.~Wooldridge, T.~Sandholm, and M.~Yokoo, editors.
\newblock {\em Proceedings of the 2nd International Joint Conference on
  Autonomous Agents and Multiagent Systems, {AAMAS} 2003, Melbourne, Australia,
  June 14--18, 2003}. International Foundation for Autonomous Agents and
  Multiagent Systems, 2003.

\bibitem[Sip82]{sip:c:complete-sets}
M.~Sipser.
\newblock On relativization and the existence of complete sets.
\newblock In {\em Proceedings of the 9th International Colloquium on Automata,
  Languages, and Programming}, pages 523--531. Springer-Verlag {\it Lecture
  Notes in Computer Science \#140}, July 1982.

\bibitem[SU02]{sch-uma:j:PH-part-one-with-web-updates-cited}
M.~Schaefer and C.~Umans.
\newblock Completeness in the polynomial-time hierarchy: Part {I}: {A}
  compendium.
\newblock {\em SIGACT News}, 33(3):32--49, 2002.
\newblock Updated version available online at {\tt
  ovid.cs.depaul.edu/documents/phcom.pdf}.

\bibitem[SV00]{spa-vog:c:theta-two-classic}
H.~Spakowski and J.~Vogel.
\newblock $ {\rm {\Theta}}_2^p$-completeness: {A} classical approach for new
  results.
\newblock In {\em Proceedings of the 20th Conference on Foundations of Software
  Technology and Theoretical Computer Science}, pages 348--360. Springer-Verlag
  {\it Lecture Notes in Computer Science \#1974}, December 2000.

\bibitem[Tod91]{tod:j:pp-ph}
S.~Toda.
\newblock {PP} is as hard as the polynomial-time hierarchy.
\newblock {\em SIAM Journal on Computing}, 20(5):865--877, 1991.

\bibitem[Val76]{val:j:checking}
L.~Valiant.
\newblock The relative complexity of checking and evaluating.
\newblock {\em Information Processing Letters}, 5(1):20--23, 1976.

\bibitem[Wag87]{wag:j:more-on-bh}
K.~Wagner.
\newblock More complicated questions about maxima and minima, and some closures
  of {NP}.
\newblock {\em Theoretical Computer Science}, 51(1--2):53--80, 1987.

\bibitem[Wag90]{wag:j:bounded}
K.~Wagner.
\newblock Bounded query classes.
\newblock {\em SIAM Journal on Computing}, 19(5):833--846, 1990.

\end{thebibliography}

\end{document}